\documentclass[showpacs,preprintnumbers,amsmath,reprint,aps,prl,superscriptaddress,twocolumn]{revtex4-2}

\usepackage{bm}
\usepackage{graphicx}
\usepackage{color}
\usepackage{amsmath}
\usepackage{hyperref}
\usepackage{epstopdf}
\usepackage{upgreek}
\usepackage{mathptmx, textcomp}
\usepackage[latin1]{inputenc}
\usepackage[T1]{fontenc}
\usepackage{array, multirow}
\usepackage[table]{xcolor}
\usepackage{booktabs}
\usepackage{longtable}

\usepackage[normalem]{ulem}

\newcommand{\commentOut}[1]{}

\newcommand{\vib}{\ensuremath{\text{v}}}

\begin{document}

\title{Spin-conservation propensity rule for three-body recombination of ultracold Rb atoms}

\author{Shinsuke Haze} \email{shinsuke.haze@uni-ulm.de}
\affiliation{Institut f\"{u}r Quantenmaterie and Center for Integrated Quantum Science and Technology IQ$^{ST}$, Universit\"{a}t Ulm, D-89069 Ulm, Germany}

\author{Jos\'{e} P. D'Incao}
\affiliation{Institut f\"{u}r Quantenmaterie and Center for Integrated Quantum Science and Technology IQ$^{ST}$, Universit\"{a}t Ulm, D-89069 Ulm, Germany}
\affiliation{JILA, NIST and Department of Physics, University of Colorado, Boulder, CO 80309-0440, USA}

\author{Dominik Dorer}
\affiliation{Institut f\"{u}r Quantenmaterie and Center for Integrated Quantum Science and Technology IQ$^{ST}$, Universit\"{a}t Ulm, D-89069 Ulm, Germany}

\author{Markus Dei{\ss}}
\affiliation{Institut f\"{u}r Quantenmaterie and Center for Integrated Quantum Science and Technology IQ$^{ST}$, Universit\"{a}t Ulm, D-89069 Ulm, Germany}

\author{Eberhard Tiemann}
\affiliation{Institut f\"ur Quantenoptik, Leibniz Universit\"at Hannover, 30167 Hannover, Germany}

\author{Paul S. Julienne}
\affiliation{Institut f\"{u}r Quantenmaterie and Center for Integrated Quantum Science and Technology IQ$^{ST}$, Universit\"{a}t Ulm, D-89069 Ulm, Germany}
\affiliation{Joint Quantum Institute, University of Maryland and NIST, College Park, MD 20742, USA}

\author{Johannes Hecker Denschlag} \email{johannes.denschlag@uni-ulm.de}
\affiliation{Institut f\"{u}r Quantenmaterie and Center for Integrated Quantum Science and Technology IQ$^{ST}$, Universit\"{a}t Ulm, D-89069 Ulm, Germany}

\date{\today}

\begin{abstract}
	We explore the physical origin and the general validity of a propensity rule for the conservation of the hyperfine spin state in three-body recombination. This rule was recently discovered for the special case of $^{87}$Rb with its nearly equal singlet and triplet scattering lengths. Here, we test the propensity rule for $^{85}$Rb for which the scattering properties are very different from $^{87}$Rb. The Rb$_2$ molecular product distribution is mapped out in a state-to-state fashion using REMPI detection schemes which fully cover all possible molecular spin states.  Interestingly, for the experimentally investigated range of binding energies from zero to $\sim13\:\textrm{GHz}\times h$ we observe that the spin-conservation propensity rule also holds for $^{85}$Rb. From these observations and a theoretical analysis we derive an understanding for the conservation of the hyperfine spin state. We identify several criteria to judge whether the propensity rule will also hold for  other elements and collision channels.  
\end{abstract}

\maketitle

The particular mechanisms of chemical reactions often give rise to selection and propensity rules. While selection rules express strict exclusion principles for product channels, propensity rules specify which product channels are more likely to be populated than others \cite{Berry1966,Fano1985}. Since the early days of quantum mechanics a central question in  reaction dynamics is whether composite spins are conserved. Wigner's spin-conservation rule, e.g., states that the total electronic spin has a propensity to be conserved \cite{Moore1973,Lee1991,Hermsmeier2021}. The recent progress in the quantum state-resolved preparation and detection of ultracold atoms and molecules has now made it possible to 
experimentally explore spin conservation rules that also involve nuclear spins. In a recent study of bimolecular reactions of ultracold KRb molecules the conservation of the total nuclear spin was found \cite{Hu2021, Liu2021}. In a study on the final state distribution of three-body recombination of ultracold $^{87}$Rb atoms our group found a propensity for the conservation of the hyperfine state of the atom pair forming the molecule \cite{Wolf2017, Wolf2019}. More precisely, this spin propensity rule states that the  angular momentum quantum numbers $F, f_a, f_b$ and $m_F = m_{fa} + m_{fb}$ are conserved in the reaction. Here, $f_a, f_b$ correspond to the total angular momenta of the two atoms ($a$, $b$) forming the molecule, and  $\vec{F} = \vec{f_a} + \vec{f_b}$.

Formally, there is no selection rule that forbids spin exchange between all three atoms in the recombination process, and a corresponding change in the  quantum numbers. In fact, recent calculations predict spin exchange to occur in three-body recombination of $^7$Li \cite{Li2021} and $^{39}$K \cite{chapurin2019PRL,xie2020PRL}, although being suppressed for $^{87}$Rb \cite{Li2021}. In order to explain the observed spin propensity rule for $^{87}$Rb one can justifiably argue that $^{87}$Rb is special since here the singlet ($a_s$) and triplet ($a_t$) scattering lengths are nearly identical ($a_s \approx 90a_0, a_t \approx 99a_0$, where $a_0$ is the Bohr radius). This leads to a strong suppression of two-atom spin exchange reactions \cite{burke1997PRA,Julienne1997,Kokkelmans1997}.

In order to explore the validity of the spin propensity rule further, we investigate here, both experimentally and theoretically, three-body recombination of ultracold $^{85}$Rb atoms which have very different two-body scattering properties from $^{87}$Rb atoms. The singlet and triplet scattering lengths for $^{85}$Rb atoms are $a_s = 2720a_0$ and $a_t = -386.9a_0$ \cite{Strauss2010}, respectively, and the three-body recombination rate constant $L_3$ for $^{85}$Rb is about four orders of magnitude larger than for $^{87}$Rb. Nevertheless, as a central result of our work, we find the spin propensity rule to also hold for $^{85}$Rb, within the investigated range of binding energies from zero to $13\:\textrm{GHz}\times h$ and the resolution of our experiment \cite{res}. This result is corroborated by the fact that our measured product distributions are well reproduced by numerical three-body calculations based on a single-spin channel. The spin propensity rule that we find for $^{87}$Rb and $^{85}$Rb will also hold for other elements if certain conditions are met, which are formulated in the present work.

\begin{figure*}[t]
	\includegraphics[width=2.07\columnwidth]{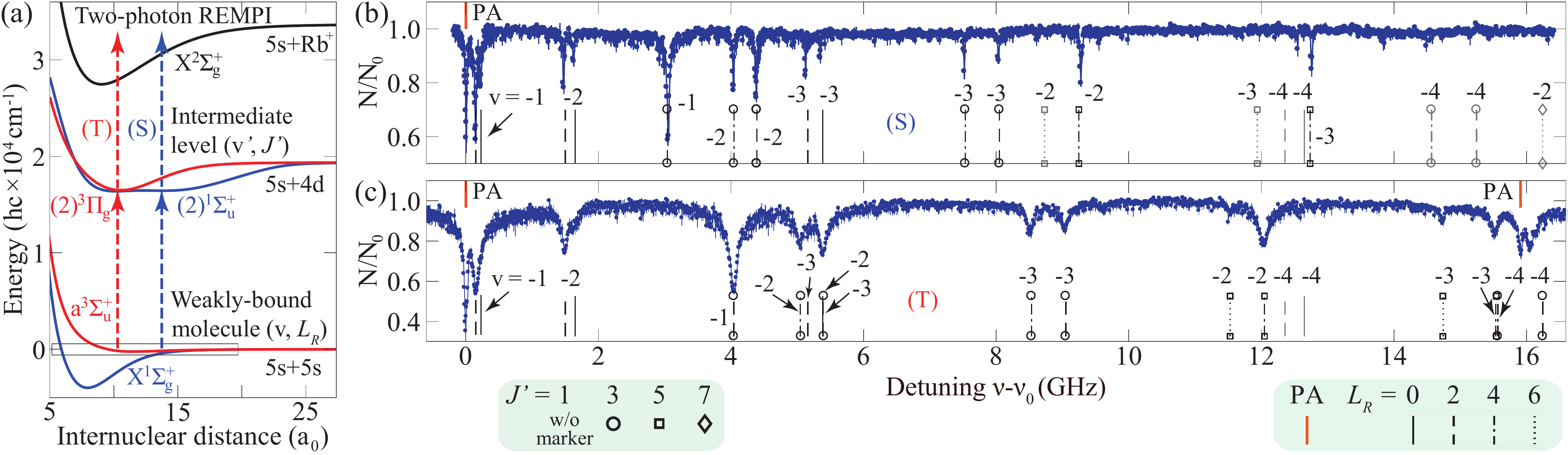}
	\caption{(a) REMPI pathways for detecting molecules with singlet $X^1\Sigma_g^+$ and/or triplet $a^3\Sigma_u^+$ character via the intermediate states $(2)^1\Sigma_u^+$, $\vib' = 22 $
		\cite{Ascoli2015} [path (S)] and $(2)^3\Pi_g$, $0_g^+$, $\vib' = 10$ \cite{Ascoli2015} [path (T)], respectively (see blue (red) dashed vertical arrows). The potential energy curves are derived from \cite{Bellos2013, Lozeille2006, Strauss2010}. (b,c) REMPI spectra of product molecules using path (S) for (b) and path (T) for (c). Shown is the normalized remaining atom number $N/N_0$ as a function of the REMPI laser frequency $\nu$, where $\nu_0=497603.591\:\text{GHz}$ (b), $\nu_0=497831.928\:\text{GHz}$ (c), are the resonance positions of the photoassociation signals (PA) of the intermediate levels with $J'=1$.	$J'$ is the  total angular momentum excluding nuclear spin. The vertical lines represent calculated resonance positions, where black (gray) color indicates experimentally observed (unobserved) states. Each line is marked with the vibrational quantum number $\vib$. $L_R$ and $J'$ are given by the linestyle and the plot symbols, respectively (see legends).}
	\label{fig3}
\end{figure*} 

We carry out the measurements with an ultracold cloud of $2.5\times 10^5$ $^{85}$Rb ground state atoms at a temperature of $860\:\text{nK}$ and at near-zero magnetic field $B$. The atoms have spin $f = 2, m_f = -2$ and are trapped in a far-detuned optical dipole trap \cite{Supp}. Three-body recombination produces weakly-bound molecules in the mixed singlet $X^1\Sigma_g^+$ and triplet $a^3\Sigma_u^+$ states [see Fig.$\:$\ref{fig3}(a)], which are coupled by hyperfine interaction \cite{Strauss2010}. We measure the yields of molecular products, observing a range of rovibrational states with vibrational and rotational  quantum numbers $\vib, L_R$, respectively,  from $\vib=-1$ to $-4$ \cite{Vib} and from $L_R = 0$ to $6$. A main result of our experiments is that we only find population in molecular states which are in the same spin state $F = 4,f_a = 2,f_b = 2$ (in short $F f_a f_b = 422$) as the reacting atom pair, although the investigated range of binding energies covers many bound states with different spin states. Because the two $^{85}$Rb atoms ($a,b$) are identical bosons, the state $F f_a f_b = 422$ goes along with only having even angular momenta $L_R$ and a positive total parity.

We have extended our previous state-selective detection scheme \cite{Wolf2017} so that it now covers all symmetries of the dimer product state space, including spin triplet and singlet states with their respective $u/g$ symmetry. We apply two-step resonance-enhanced multiphoton ionization (REMPI), similar to \cite{Lozeille2006, Huang2006, Gabbanini2000, Jyothi2016, Mancini2004}, but with a cw-laser. By two different REMPI pathways, (S) or (T), we probe product molecules via singlet or triplet character [see Fig.$\:$\ref{fig3}(a)]. Both schemes use identical photons for the two REMPI steps at wavelengths around $602.2\:\text{nm}$ (see  \cite{Supp}). The intermediate states are deeply-bound levels of 
$(2)^1\Sigma_u^+$ and $(2)^3\Pi_g$ for REMPI (S) and (T), respectively. When ions are produced via REMPI, they are directly trapped and detected in an eV-deep Paul trap which is centered on the atom cloud. Subsequently, elastic atom-ion collisions inflict tell-tale atom loss while the ions remain trapped. From the atom loss which is measured via absorption imaging of the atom cloud \cite{Haerter2013, Haerter2013b, Wolf2017} the ion number can be inferred \cite{Supp}.

Figures \ref{fig3}(b) and (c) show REMPI spectra of Rb$_2$ product molecules following three-body recombination, using path (S) and (T), respectively. Apart from three signals stemming from the photoassociation of two atoms (indicated by PA) \cite{Jones2006}, each resonance line of loss corresponds to a molecular product state. The photoassociation lines serve as references for the $|f=2,m_f=-2 \rangle
+ |f=2,m_f=-2 \rangle$ asymptote corresponding to zero binding energy at about zero magnetic field. The vertical lines in Figs.$\:$\ref{fig3}(b) and (c) are predicted frequency positions for product states for  the spin state $Ff_af_b=422$. These predictions are obtained from  coupled-channel calculations for the $X^1\Sigma_g^+-a^3\Sigma_u^+$ complex \cite{Strauss2010}. Coincidences of predicted and observed lines allow for an assignment. As a consistency check for the assignment of signals in Figs.$\:$\ref{fig3}(b) and (c) we make use of  product states with $L_R>0$ since these give rise to two resonance lines corresponding to the transitions towards $J'=L_R\pm1$. Indeed, the data in Figs.$\:$\ref{fig3}(b) and (c) confirm this consistency. Inspection clearly shows, that all experimentally observed spectral lines in Figs.$\:$\ref{fig3}(b) and (c) can be explained as signals from product molecules with the spin $F f_a f_b = 422$. As an additional check for the line assignment, we show in \cite{Supp} that our experimental spectra do not match up with molecular spin states other than $F f_a f_b = 422$. This clearly indicates that the same spin propensity rule previously observed for $^{87}$Rb also holds for $^{85}$Rb. 

For each product signal in the  singlet REMPI path (S) we obtain a corresponding signal  in the triplet REMPI path (T). This is because the spin state $F f_a f_b = 422$, $m_F = -4$ has sizeable singlet ($\approx15\%$) and triplet ($\approx85\%$) admixtures. The spectra of Figs.$\:$\ref{fig3}(b) and (c) generally look different since the rotational constants differ 
for $(2)^1\Sigma_u^+$  ($B_{\vib'} = 289(1)\:$MHz) and for $ (2)^3\Pi_g, 0_g^+$
($B_{\vib'} = 389(2)\:$MHz), see, e.g., \cite{Guan2013}. The linewidths in Fig.$\:$\ref{fig3}(c)
are typically on the order of $100\:\text{MHz}$ (FWHM) which is larger than the typical linewidths in Fig.$\:$\ref{fig3}(b) of about $30\:\text{MHz}$. This is a consequence of the larger hyperfine splitting of the $(2)^3\Pi_g$ state, which is not resolved in our measurements. In Fig.$\:$\ref{fig3}(c) at $\nu-\nu_0=15.90\:\text{GHz}$ there is a photoassociation signal which belongs to the $0_g^-$ component of $(2)^3\Pi_g$. We show the corresponding REMPI spectrum in \cite{Supp}. It exhibits the same molecular states as in Figs.$\:$\ref{fig3}(b) and (c).

We now carry out a more quantitative analysis where we compare the experimental signal strengths of the various REMPI paths and also compare them to theoretical calculations. For this, we measure the ion production rate for each assigned resonance line in  Fig.$\:$\ref{fig3} and Fig.$\:$S1 \cite{Supp}. For a given REMPI path, the ion rate signal is expected to be proportional to the product molecule population rate, if we assume equal REMPI ionization efficiencies for the states. Our data show that this assumption is indeed fulfilled for most data points within the uncertainty limits of the recordings.

\begin{figure}[t]
	\includegraphics[width=\columnwidth]{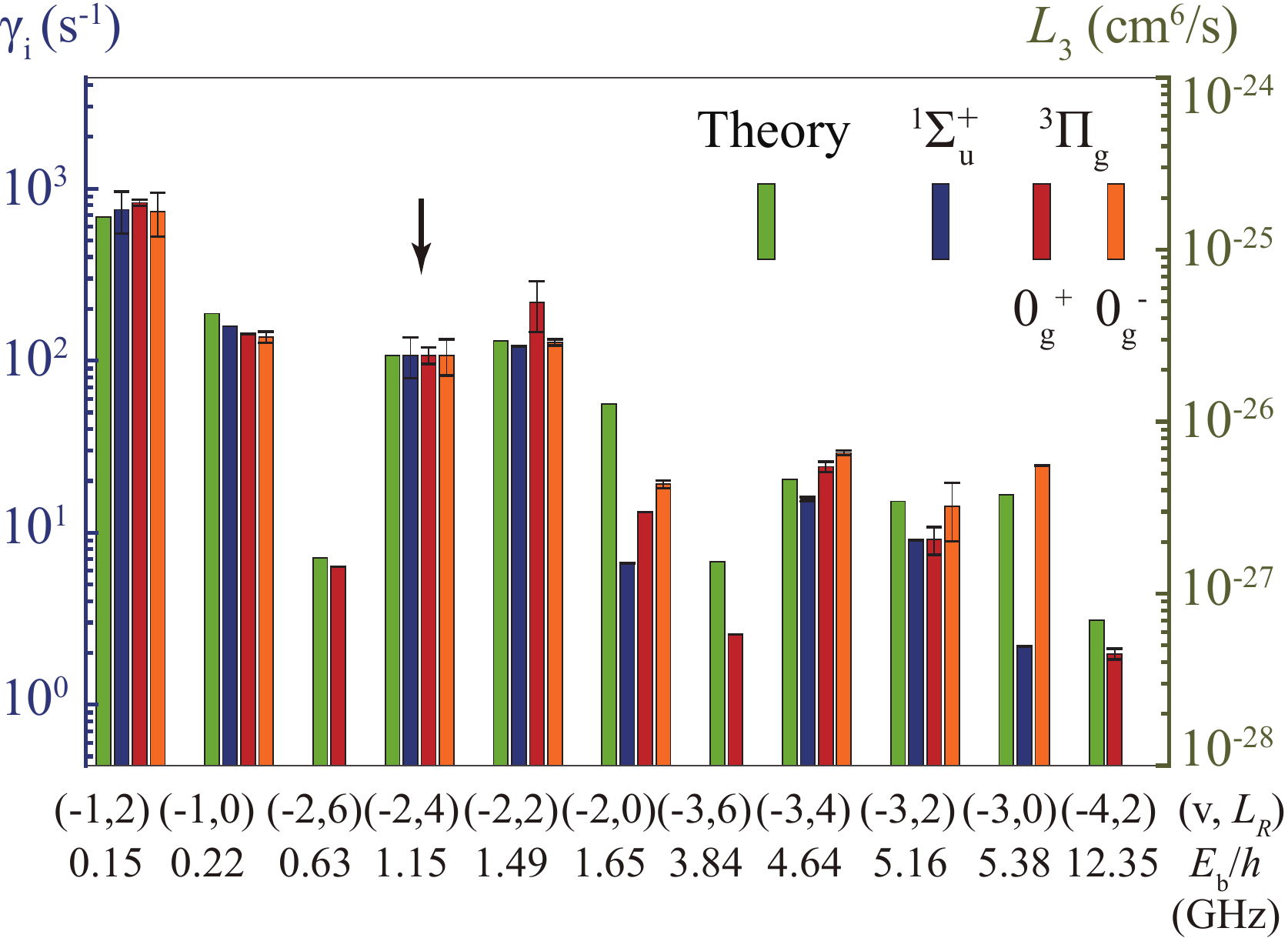}
	\caption{Comparison of calculations and experiments. Measured (scaled) ion production rates $\gamma_i$ for the product states $(\vib, L_R)$ with binding energies $E_b$ are given for the three REMPI paths  together with calculations for the three-body recombination channel rate constants $L_3(\vib,L_R)$ (see the legend for the color coding). The ion signals for the  $(2)^3\Pi_g$ $0_g^-$ and $(2)^1\Sigma_u^+$ paths have been scaled (see text).}
	\label{fig_histo}
\end{figure}

Figure$\:$\ref{fig_histo} shows the extracted ion production rates $\gamma_i$ of each molecular product state for the three REMPI paths (via $(2)^1\Sigma_u^+$,  $(2)^3\Pi_g$ $0_g^+$, and  $(2)^3\Pi_g$ $0_g^-$). If a state is observed via two or three different $J'$ levels for a given path, we plot the average of the rates and mark the standard deviation from the mean with an error bar. In order to ease the comparison between the three data sets we have globally scaled the ion signals for the $(2)^3\Pi_g$ $0_g^-$ and   $(2)^1\Sigma_u^+$ paths by a factor of 7.3 and 3.2, respectively, so that the signal bars for all REMPI paths in Fig.$\:$\ref{fig_histo} are the same height for the state ($\vib=-2$, $L_R = 4$), see black arrow. These scaling factors compensate the differences in the ionization efficiencies of the different REMPI paths, which are due to the different singlet and triplet components of the product molecule as well as differences in the electric dipole transition moments, which are generally not very well known yet \cite{uncert}. After the scaling the signals of the three paths for a given bound state are consistent over the set of detected product states.

In addition to the experimental data, we plot in Fig.$\:$\ref{fig_histo} calculated channel rate constants $L_3(\vib,L_R)$ for a temperature of 0.8$\:\mu$K. The calculations use a single-spin model \cite{Supp} which is similar to the one used in Ref.$\:$\cite{Wolf2017} in order to solve the three-body Schr{\"o}dinger equation in an adiabatic hyperspherical representation \cite{dincao2018JPB,wang2011PRA}. For this, we use pairwise additive long range van der Waals potentials with a scattering length of $-$443a$_0$ \cite{claussen2003PRA} for $^{85}$Rb and with a truncated number $Z$ of $L_R=0$ bound states ($Z=9$ for Fig.$\:$\ref{fig_histo}). The calculated total recombination rate constant at 0.8$\:\mu$K (including thermal averaging) is $L_3 = 3.07\times 10^{-25}$ cm$^6/$s and is consistent with the values found in Ref.$\:$\cite{Roberts2000}.

All data sets in Fig.$\:$\ref{fig_histo} display the trend that the population of a molecular state due to three-body recombination typically decreases with increasing  binding energy $E_b$ of the product state, which is consistent with the work for $^{87}$Rb \cite{Wolf2017}. The overall agreement between theory and experiment in Fig.$\:$\ref{fig_histo} is good, as the experimentally observed relative strengths of the signals for low $L_R$ are in general well reproduced by the calculated recombination rates. This suggests that our single-spin model fully captures the characteristics of the three-body chemical reaction in the given parameter regime, which can be viewed as additional evidence for the spin propensity rule. 

Based on our theoretical analysis, we conclude that the spin propensity rule in our experiment is a consequence of the following scenario. a) The reaction takes place at interparticle distances where the interactions of particles $a$ and $b$ (forming the molecule) with particle $c$ are nearly spin-independent. b) In the investigated range all possible product molecules have quantum states with good quantum numbers $F, f_a, f_b$. c) In the reaction region, the spin composition of the reacting pair $a, b$ is essentially given by $F  f_a f_b=422$. As a consequence of conditions a), b) and c) the molecule formation is driven by mechanical forces from atom $c$ while the spin state of the reacting pair is not affected. 

To show that this scenario holds for our experiments, we first analyze the typical interparticle distances where the reaction occurs. Our numerical calculations \cite{Supp} show that the formation of $^{85}$Rb$_2$ molecules mainly takes place near a hyperradius $R \approx 1.5 r_{\rm vdW}$, extending from $R \approx 1.1r_{\rm vdW}$ to 2$ r_{\rm vdW}$. Here, $r_{\rm vdW} = ( 2\mu C_6 / \hbar^2)^{1/4} = 82a_0$ denotes the van der Waals length for Rb, and $\mu$ and $C_6$ are the reduced mass and the van der Waals coefficient of the two-particle system, respectively. The hyperradius $R$ describes the characteristic size of the three-body system and is given by $R^2 = (\vec{r}_b - \vec{r}_a )^2 /d^2 + d^2 (\vec{r}_c - (\vec{r}_a + \vec{r}_b )/2  )^2 $, where $\vec{r}_i$ is the location of particle $i$ and $d^2 = 2/\sqrt{3}$ \cite{dincao2018JPB}. The fact that the reactions occur at these large $R$ can be understood within the framework of the adiabatic hyperspherical representation, where an effective repulsive barrier for the three-body entrance channel forms at a hyperradius of about $R = 1.7 r_{\rm vdW}$ \cite{Wang2012prl,Supp}.

We now consider the formation of a molecule state with a size $\lesssim r_{\rm hf}\approx0.6r_{\rm vdW}$. Only for such a state can spin components other than $F f_a f_b = 422$ be substantial (for more details see \cite{Supp}). Here, $r_{\rm hf}=(C_6/E_{\rm hf})^{1/6}$ is the hyperfine radius and $E_{\rm hf} = 3.04\:\textrm{GHz}\times h$ is the atomic hyperfine splitting. Given the hyperradius $R > 1.1 r_{\rm vdW} $ and the interparticle distance
$r_{ab}  \equiv  | \vec{r}_a - \vec{r}_b | < r_{\rm hf}$, the distances of particle $c$ to the others must be $r_{ca}$, $r_{cb} >  0.6r_{\rm vdW}$. Since the interaction between
two Rb atoms is essentially spin-independent for distances  $\gtrsim  0.25 r_{\rm vdW}$ \cite{Supp}, this validates point a) of the scenario described above. Concerning point b), our coupled-channel calculations show that the weakly-bound molecular states up to a binding energy of about $50\:\mathrm{GHz}\times h$ have almost pure spin states $F f_a f_b $. This can be explained by the fact that the $^{85}$Rb triplet and singlet scattering lengths are large in magnitude. For $^{85}$Rb, where these scattering lengths are of opposite sign, this leads to a near energetical degeneracy of the triplet vibrational levels $\vib_T$ with the singlet vibrational levels $\vib_S = \vib_T - 1$ \cite{Julienne2009, Julienne2009b}. Since, in addition, the singlet and triplet vibrational wavefunctions are very similar at long range, the interaction between the  two atoms is effectively spin-independent. As a consequence the atomic hyperfine interaction of each atom is essentially unperturbed, which leads to the atomic hyperfine structure with the quantum numbers $f_a, f_b$ for the molecule. Subsequent coupling of $\vec{f_a}, \vec{f_b}$ in the molecule forms a total $\vec{F}$.

To show point c), we first note that due to the magnitude of $R$ the three-body system can be effectively decomposed (with respect to spin)  into a two-body collision of two $f = 2, m_f = -2$ atoms ($a$ and $b$) and a third atom ($c$) which is spin-wise only a spectator, see also \cite{Supp}. During the two-body collision of atoms $a$ and $b$, spin admixtures to the original 422 state can occur at close distance. However, these admixtures are only on the \% level for $^{85}$Rb even for short atomic distances $r_{ij} \sim 0.25 r_{\rm vdW}$, and therefore negligible. That this admixture is small can be derived from the fact, that the two-body scattering wavefunction is very similar to the one of the corresponding last molecular bound state at short distance. The scattering state therefore shares the relatively pure $F f_a f_b $ spin state character of the weakly-bound molecular states, as discussed before.

\begin{figure}[t]
	\includegraphics[width=\columnwidth]{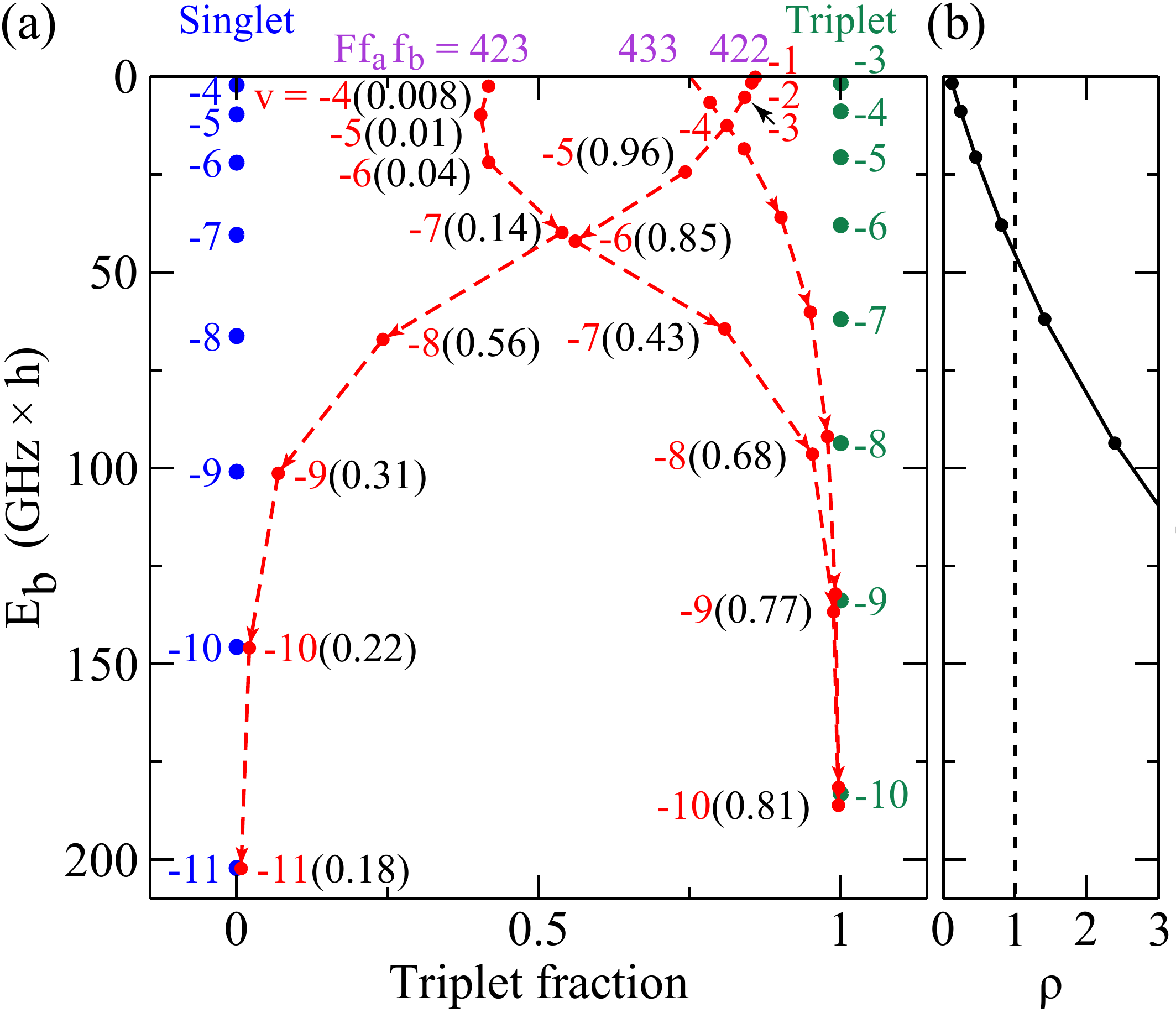}
	\caption{(a) Binding energy and spin character of weakly-bound $^{85}$Rb$_2$ molecules for various vibrational quantum numbers $\vib = -1$ to $-11$ at $B=0$ (red circles). Here, $L_R=0$, $F = 4$, and $m_F = -4$. Zero energy corresponds to two separated atoms in state $f_a f_b = 2 2$. The spin character is given as the norm of the spin triplet component. Shown are the three spin families with $F = 4$ which correlate with the states $Ff_af_b = 422, 423, 433$ at the $E_b = 0$ threshold. Red dashed arrows are guides to the eye and indicate the change of each family's spin character with increasing binding energy. The norm of the 422 spin component is presented in parentheses for the 423 and 422 families. For the 433 family the 422 norm is about 1\% for the shown bound states. Blue and green circles are the bound state levels of the pure singlet and triplet potentials, respectively, the dissociation limit of which is at $E=+3.542\:\textrm{GHz}\times h$ above $E_b=0$. (b) Ratio $\rho$ for various binding energies (see text).}
	\label{Schematic}
\end{figure}

Our coupled-channel calculations using the potentials of \cite{Strauss2010} show that for molecular bound states with a binding energy larger than $50\:\text{GHz}\times h$ the spin decomposition changes, because the splitting of adjacent singlet and triplet vibrational levels becomes larger than the hyperfine splitting. This is shown in Fig.$\:$\ref{Schematic}(a). In our experiment we only observe product states down to the $\vib=-4$ level in the 422 family, for which the norm of the 422 entrance channel remains between 0.99 and unity. All other unobserved spin families have a norm of the 422 channel much less than unity. In Fig.$\:$\ref{Schematic}(b) we plot the ratio $\rho$ of the triplet and singlet level splitting and the hyperfine splitting, $\rho = |E_\mathrm{T}(\vib)-E_\mathrm{S}(\vib-1)| / E_{\mathrm{hf}} $. For $\rho > 1$ hyperfine mixing is increasingly suppressed and the propensity rule is expected to break down. 

Similar arguments can be used to also explain the  spin propensity rule for $^{87}$Rb as observed in \cite{Wolf2017, Wolf2019}. Here, the singlet and triplet scattering lengths are almost equal, leading to a near degeneracy of the singlet and triplet vibrational states for $\vib_T=\vib_S$. Furthermore, the atomic hyperfine splitting in $^{87}$Rb is larger by a factor of $\approx 2.4$ than in $^{85}$Rb. As a consequence the spin propensity rule holds for even larger binding energies of up to $100\:\text{GHz}\times h$ (see Fig$\:$S5 in \cite{Supp}). 

In the future, it will be interesting to investigate the breakdown of the spin propensity rule for $^{85}$Rb by studying product molecules with binding energies larger than $\sim 50\:\text{GHz}\times h$. For these measurements the ability to state-selectively detect singlet and triplet product molecules as demonstrated here, will be essential. The spin propensity rule will also break down when $F$ is not a good quantum number anymore, e.g., by applying strong magnetic fields. $^{85}$Rb features a broad Feshbach resonance at 153.3$\:$G	\cite{Blackley2013} where spin-mixing in the incoming channel naturally becomes important \cite{chapurin2019PRL,xie2020PRL,secker2021PRA}. Besides for Rb the spin-conservation propensity rule may hold for other elements. Cs, e.g., might be a good candidate when working in a regime where dipolar relaxation processes are negligible.

\begin{acknowledgements}  
	This work was financed by the Baden-W\"urttemberg Stiftung through the Internationale
	Spitzenforschung program (contract BWST ISF2017-061) and by the German Research Foundation (DFG, Deutsche Forschungsgemeinschaft) within contract 399903135. J.~P.~D. also acknowledges partial support from the U.S. National Science Foundation, Grant No. PHY-2012125, and NASA/JPL 1502690. The authors would like to thank Jinglun Li for helpful discussions. J.~P.~D. thanks Timur Tscherbul for stimulating discussions.
\end{acknowledgements}


\begin{thebibliography}{52}
\bibitem [{Vib()}]{Vib}%
	\BibitemOpen
	\href@noop {} {\bibinfo {title} {Here, $\vib$ is counted downwards, starting
			with $-1$ for the last $s$-wave bound state just below the atom pair
			asymptote $|f=2,m_f=-2\rangle +|f=2,m_f=-2\rangle$.}}\BibitemShut {Stop}%
\bibitem [{unc()}]{uncert}%
	\BibitemOpen
	\href@noop {} {\bibinfo {title} {We note that because of these
			uncertainties and uncertainties in the atom number calibration we refrain
			from converting the measured ion production rates into three-body
			recombination rate constants for each product channel.}}\BibitemShut {Stop}%
\end{thebibliography}
%

\section {Supplemental Material}

\subsection{Ultracold atoms set-up} 

The ultracold $^{85}$Rb atoms are initially prepared in the spin state $f = 2, m_f = -2$ of the electronic ground state. This state is stable in two-body collisions due to energetic closure of other atomic exit channels and due to small dipolar relaxation rates. Indeed, our coupled-channel calculations have verified that the two-body spin-relaxation rate constants for
$^{85}$Rb remain below $10^{-14}\:\textrm{cm}^3/$s when the magnetic field is not near a Feshbach resonance. The scattering length for the collision of two $f = 2, m_f = -2$ atoms is $a=-460a_0$.

In our set-up the atoms are trapped in a far-detuned optical dipole trap with trapping frequencies $\omega_{x,y,z}=2\pi\times (156,\,148,\,18)\:\text{Hz}$. The optical dipole trap operates at a wavelength of about $1064\:\text{nm}$.

\subsection{REMPI}

For REMPI we use a laser wavelength around $602.2\:\textrm{nm}$. The laser light is either provided by a cw dye laser (Matisse, Sirah Lasertechnik GmbH) or an optical parametric oscillator (C-Wave, H\"ubner GmbH). Each laser is stabilized to a cavity and has a short-term linewidth of less than $1\:\text{MHz}$. Longer term drifts are compensated by a lock to a wavelength meter and we have a shot-to-shot frequency stability on the order of $\pm5\:\text{MHz}$. The laser beam has a power of $100\:\text{mW}$ and a beam waist ($1/e^2$ radius) of $1\:\text{mm}$ at the location of the molecules. It has a mixture of $\sigma$- and $\pi$-polarization and can be essentially considered as unpolarized.

REMPI produces Rb$^+_2$ molecules with binding energies larger than $479\:\text{cm}^{-1}$ in the $X^2\Sigma_g^+$ state. We have no indication of a resonance structure for the ionization transition, in agreement with \cite{Gabbanini2000, Pichler1983} where the ionization range is referred to as a diffuse band.

Coincidences of predicted and observed lines in REMPI spectra allow for an assignment of the observed lines. The precision of coincidences was typically around 10$\:$MHz, as mainly determined by slight drifts of the wavemeter.

We note that besides the triplet and singlet paths discussed in detail in this publication 
we also have tested a {REMPI} path via the ${A}^1{\Sigma}_u^+$ state (associated with the $5s+5p$ asymptote) similar to the one described in \cite{Wolf2017}, however, the overall detection efficiencies are much lower in that case. Furthermore, the spectra of Figs.$\:$1(b) and (c) and Fig.$\:$\ref{fig6} are very clean in a sense that essentially no unidentified resonance signals are visible which represents a significant improvement as compared to \cite{Wolf2017}.

\subsection{The $(2)^3\Pi_g$, $0_g^-$ intermediate state}

\begin{figure*}[t]
	\renewcommand{\thefigure}{S1}
	\includegraphics[width=2.07\columnwidth]{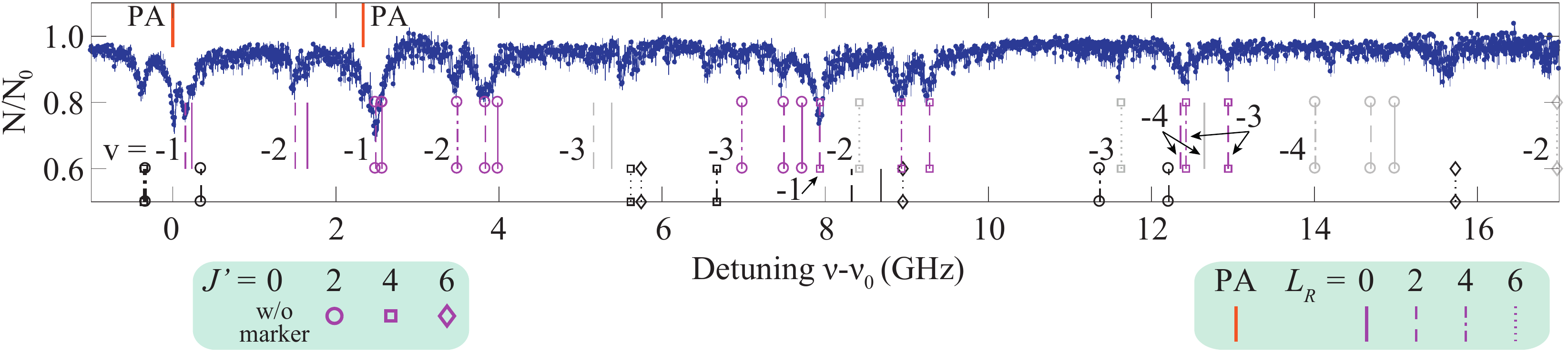}
	\caption{REMPI spectrum via the triplet path (T) with intermediate state $(2)^3\Pi_g$, $0_g^-$, $\vib'=10$. The remaining atom fraction $N/N_0$ is plotted as a function of the REMPI laser frequency $\nu$. $\nu_0=497847.828\:\text{GHz}$ is the resonance frequency for  photoassociation (PA) towards $(2)^3\Pi_g$, $0_g^-$ ($J'=0$). A second photoassociation line for a transition towards $(2)^3\Pi_g$, $0_g^-$ ($J'=2$) is located at around $\nu - \nu_0 = 2.4\:$GHz. The vertical lines are predicted possible line positions of molecular states. If observed they are purple ($0_g^-$) or black ($0_g^+$), if unobserved they are gray. The lines involving $0_g^-$ are marked with the vibrational quantum number $\vib$. The rotational quantum number $L_R$ and the intermediate level $J'$ are given by the linestyle and the plot symbols, respectively (see legends).}
	\label{fig6}
\end{figure*}

Figure \ref{fig6} shows the REMPI spectrum using the intermediate state $(2)^3\Pi_g$, $0_g^-$.
For $0_g^-$ the total parity is given by $(-1)^{J' + 1}$. Therefore, since the initial product molecules have positive total parity, only transitions towards intermediate states with even quantum number $J'$ are possible. Concerning photoassociation we now observe two lines, one for $J' = 0$ and the other one for $J' = 2$. The vertical lines are calculations for the expected REMPI signals, where we use the same rotational constant of $B_{\vib'}=389(2)\:\text{MHz}$ for the intermediate state as for $(2)^3\Pi_g$, $0_g^+$. The calculated resonance positions agree well with the measurements. As a consistency check for the assignment of signals in Fig.$\:$\ref{fig6} we use that product states can give rise to up to three resonance lines corresponding to the transitions towards $J'=L_R$ and $J'=|L_R\pm2|$. The data in Fig.$\:$\ref{fig6} confirm this consistency.

\subsection {Ion number calibration}

\begin{figure}[t]
	\renewcommand{\thefigure}{S2}
	\includegraphics[width=\columnwidth]{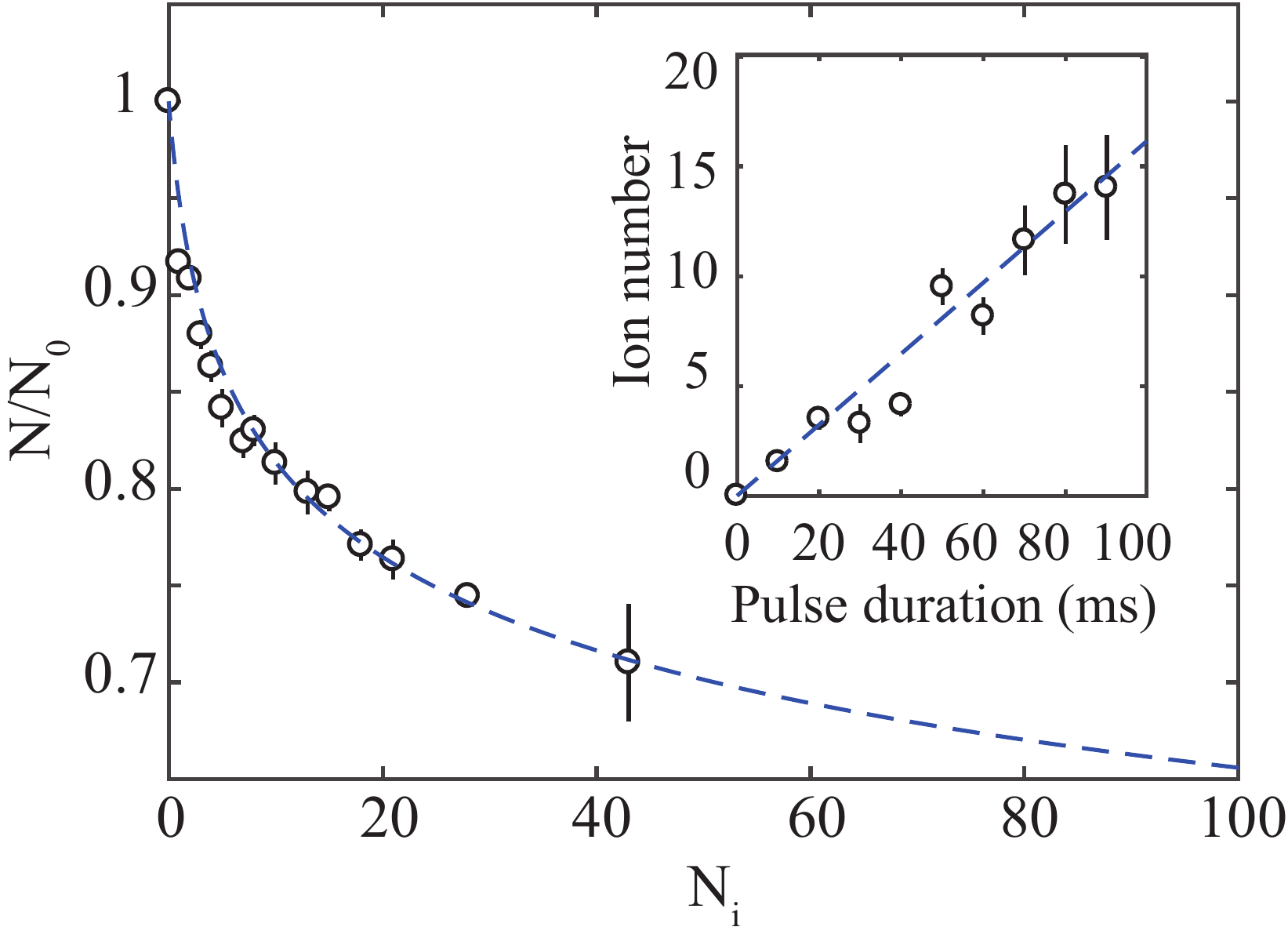}
	\caption{Atom loss due to elastic collisions with $N_i$ ions (predominantly Rb$^{+}$). Shown is the remaining atom fraction $N/N_0$ of the atom cloud after 500$\:$ms of atom-ion collisions with $N_i$ ions previously prepared in the Paul trap. The dashed line is the fit function $N/N_0 = (\frac{N_\text{i}+\alpha}{\alpha})^{\beta}$ with the fit parameters $\alpha=1.4$ and $\beta=-10.1$. The inset is a typical measurement of the ion number as a function of the REMPI laser pulse duration. For the given example, the REMPI laser frequency is set to resonantly excite  the molecular state $(\vib,L_R) = (-1,2)$ towards the $(2)^3\Pi_g$, $0_g^-$, $\vib'=10$, $J'=2$ intermediate state. The dashed line is a linear fit to the data.}
	\label{fig S1}
\end{figure}

\begin{figure}[t]
	\renewcommand{\thefigure}{S3}
	\includegraphics[width=\columnwidth]{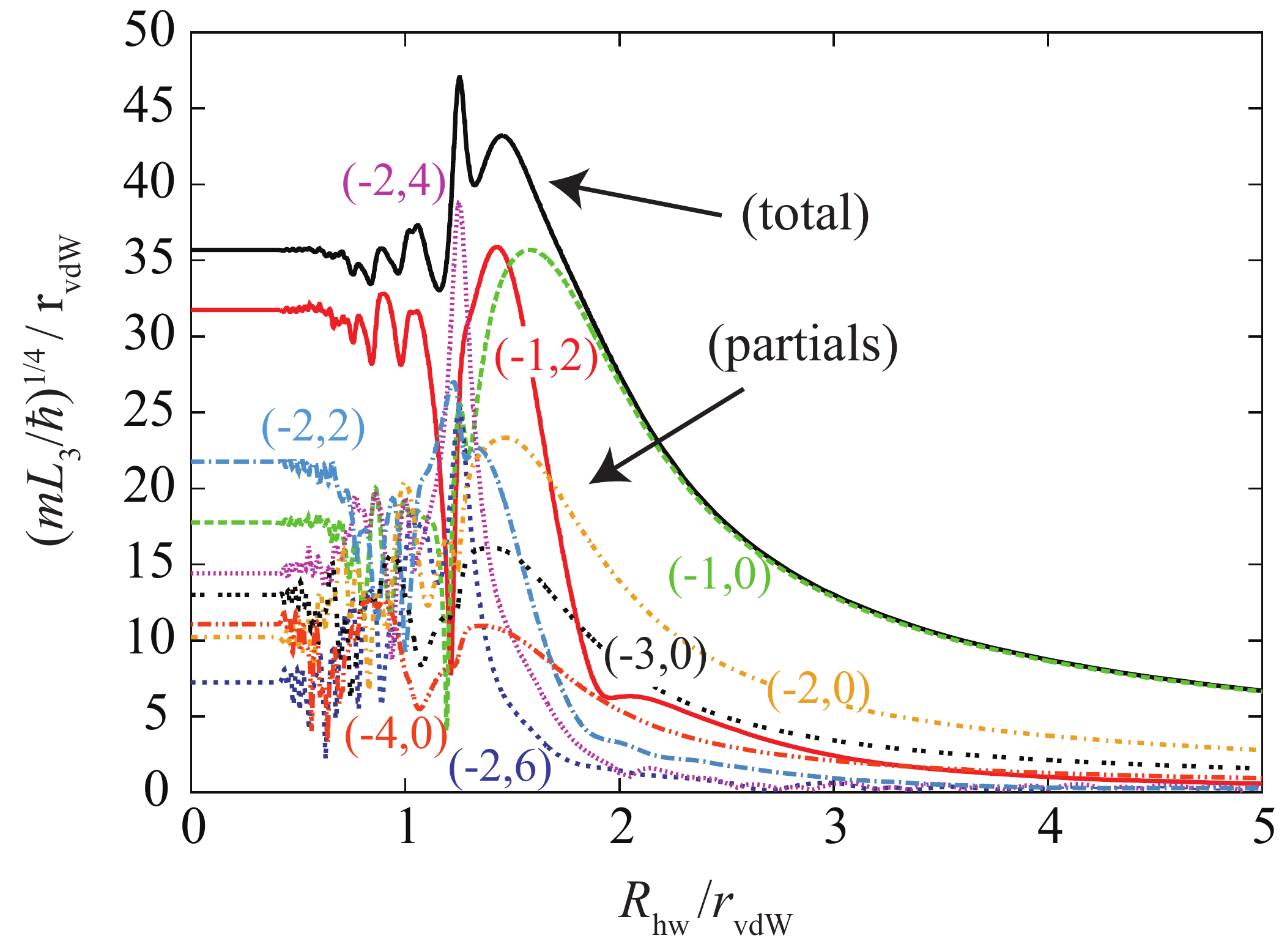}
	\caption{Total three-body recombination rate $L_{3}$ (black solid line) and corresponding partial rates for the six most weakly-bound molecular states currently observed, $(\vib,L_R)$ = (-1,2), (-1,0), (-2,6), (-2,4), (-2,2) and (-2,0), as well as (-3,0) and (-4,0), calculated with a two-body interaction model supporting 5 $s$-wave bound states. These results indicate that recombination is likely to occur at $R\approx1.5r_{\rm vdW}$, while it is increasingly suppressed for $R<1.5r_{\rm vdW}$ likely due to the presence of a repulsive barrier on the income collision channel (see text) and for $R>1.5r_{\rm vdW}$ due to the decrease of the hyperradial inelastic couplings with $R$ \cite{dincao2005PRA}.}
	\label{L3Rhw}
\end{figure}

For each molecular resonance line in the spectra of Fig.$\:$1 and Fig.$\:$\ref{fig6} we have measured the ion production rate in our experiment. For this, we tune the ionization laser onto a given resonance and  turn it on for a short enough pulse time so that only a few ions ($<20$) are produced. We are then in a regime where the ion number grows linearly with time (see inset of Fig.$\:$\ref{fig S1} for a typical measurement). Afterwards we count the number of ions by inserting them into a fresh atom cloud with $2.5\times 10 ^{5}$ atoms. After an interaction time of 500$\:$ms, during which the ions inflict atom loss due to  (mostly)  elastic atom-ion collisions \cite{Haerter2013b}, we measure the remaining number of atoms with absorption imaging. Using the calibration curve in Fig.$\:$\ref{fig S1} we can convert the measured atom fraction into an ion number. Dividing the ion number by the pulse time we obtain the ion production rate.

The calibration curve was obtained as follows. Initially, the ion trap and the optical dipole trap are spatially separated from each other so that the atoms cannot collide with the ions.  A known number of laser-cooled $^{138}$Ba$^+$ ions in a range between 1 to about 40 is prepared in the Paul trap. The cold ions form an ion crystal in the trap and are counted  with single-particle resolution after fluorescence imaging. In parallel, we prepare a $^{85}$Rb atom cloud in the crossed optical dipole trap. Next, the centers of the ion and atom traps are overlaid, immersing the Ba$^+$ ions into the atom cloud. Subsequently, the  Ba$^+$ ions can undergo reactions  with the Rb atoms, such as, e.g., charge exchange \cite{Mohammadi2021}. After a long interaction time of 2s we are predominantly left with Rb$^+$ ions, as confirmed by mass spectrometry. Although these ions can have high kinetic energy they are  still confined in the eV-deep Paul trap, i.e., the initial number of trapped ions is conserved. Afterwards, the ion trap and the optical dipole trap are separated again from each other, the old atom cloud is discarded and a new atom cloud (with 2.5$\times 10 ^{5}$ atoms) is prepared. The ions are immersed into the new atom cloud for $500\:\textrm{ms}$  and atom-ion collisions inflict again atom loss. Figure \ref{fig S1} shows the remaining atom fraction $N/N_0$ as a function of initial ion number $N_i$. The dashed line is a fit with the empirically-found  function $N/N_0 = (\frac{N_\text{i}}{\alpha}+1)^{\beta}$. Here,  $\alpha$ and $\beta$ are fit coefficients.

\subsection{Three-body model for $^{85}$Rb atoms}

Our three-body calculations for $^{85}$Rb atoms were performed using the adiabatic hyperspherical representation \cite{dincao2018JPB,wang2011PRA} where the hyperradius $R$ determines the overall size of the system, while all other degrees of freedom are represented by a set of hyperangles $\Omega$. Within this framework, the three-body adiabatic potentials $U_{\alpha}$ and channel functions $\Phi_{\alpha}$ are determined from the solutions of the hyperangular adiabatic equation:
\begin{align}
	\left[\frac{\Lambda^2(\Omega)+15/4}{2\mu R^2}\hbar^2+\sum_{i<j}v(r_{ij})\right]\Phi_{\alpha}(R;\Omega)
	=U_{\alpha}(R)\Phi_{\alpha}(R;\Omega),
\end{align}
which contains the hyperangular part of the kinetic energy, expressed through the grand-angular momentum operator $\Lambda^2$ and the three-body reduced mass $\mu=m/\sqrt{3}$. To calculate the three-body recombination rate we solve the hyperradial Schr\"odinger equation \cite{wang2011PRA},
\begin{align}
	&\left[-\frac{\hbar^2}{2\mu}\frac{d^2}{dR^2}+U_{\alpha}(R)\right]F_{\alpha}(R)\nonumber\\
	&~~~~~~+\sum_{\alpha'}W_{\alpha\alpha'}(R)F_{\alpha'}(R)=EF_{\alpha}(R),\label{Schro}
\end{align}
where $\alpha$ is an index that labels all necessary quantum numbers to characterize each channel, and $E$ is the total energy. From Eq.$\:$(\ref{Schro}) we determine the $S$-matrix and the recombination rate $L_3$ \cite{wang2011PRA}.

In this present study, the interaction between $^{85}$Rb atoms is modeled by a potential similar to that used in Ref.$\:$\cite{Wolf2017}, and given by a modified Lenard-Jones potential,
\begin{align}
	v(r)=-\frac{C_6}{r^6}\left(1-\frac{\lambda^6}{r^6}\right)-\left(\frac{C_{8}}{r^8}+
	\frac{C_{10}}{r^{10}}[f_{\lambda}(r)]^{2}\right)[f_{\lambda}(r)]^{12},\label{vpot}
\end{align}
where $C_6=4710.431E_h$a$_0^6$, $C_8=576722.7E_h$a$_0^8$, and $C_{10}=75916271E_h$a$_0^{10}$, are the van der Waals dispersion coefficients from Ref.~\cite{Strauss2010}. Here, $f_{\lambda}(r)=\tanh(2r/\lambda)$ is a cut-off function that suppresses the divergence of the $C_8$ and $C_{10}$ interaction terms for vanishing interatomic distance $r$. We can adjust the value of $\lambda$ to have different numbers of diatomic bound states supported by the interaction, while still reproducing the value of the background scattering length
$a_{\rm bg}=-443a_0$ \cite{claussen2003PRA}. Our calculations were performed using $\lambda=20.72577414766491a_0$, producing 9 $s$-wave ($L_R=0$) bound states, and a total of 101 bound states including higher partial-wave states, $L_R > 0$. Our numerical calculations for three-body recombination through the solutions of Eq.~(\ref{Schro}) have included up to 200 hyperspherical channels leading to a total rate converged within a few percent. The calculated total recombination rate constant at 0.8$\:\mu$K (including thermal averaging) is $L_3 = 3.07\times 10^{-25}$ cm$^6/$s and is consistent with the values found in  Ref.$\:$\cite{Roberts2000}.

\begin{figure}[t]
	\renewcommand{\thefigure}{S4}
	\includegraphics[width=3in]{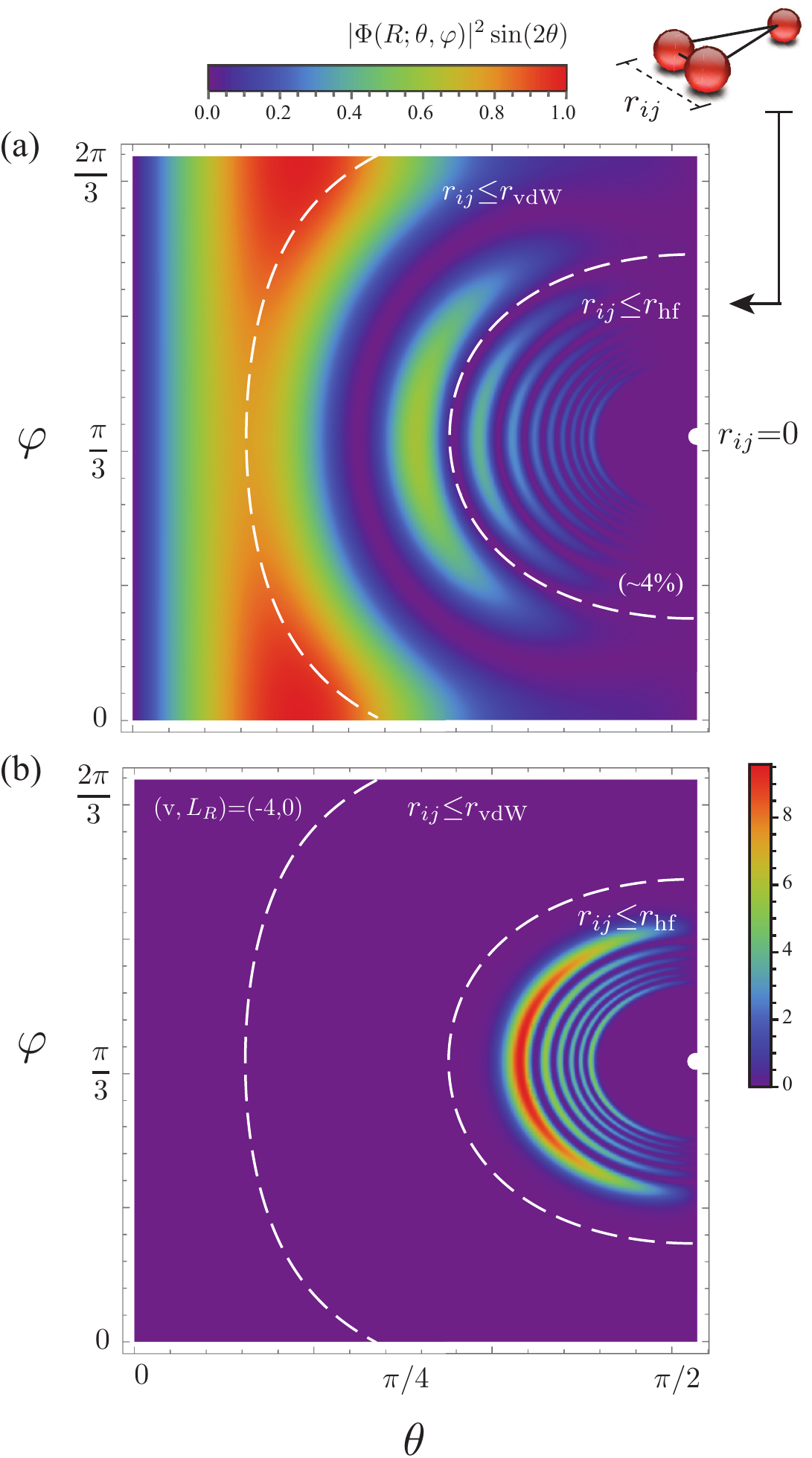}
	\caption{Hyperangular probability density, $|\Phi_{\alpha}(R;\theta,\varphi)|^2\sin(2\theta)$ \cite{Wang2012prl}, at $R=1.593r_{\rm vdW}$ for the initial collision channel (a) and the molecular product state $(\vib,L_R)=(-4,0)$ (b). A point on the $\theta$-$\varphi$ hyperangular plane specifies the geometry of the three-atom system. The regions within the circles marked by dashed lines, correspond to geometries for which two of the atoms are found at distances $r_{ij} \le r_{\rm vdW}$ and $r_{ij} \le r_{\rm hf}$, respectively, while the third atom is further away.}
	\label{ChanFun}
\end{figure}

\begin{figure}[t]
	\renewcommand{\thefigure}{S5}
	\includegraphics[width=\columnwidth]{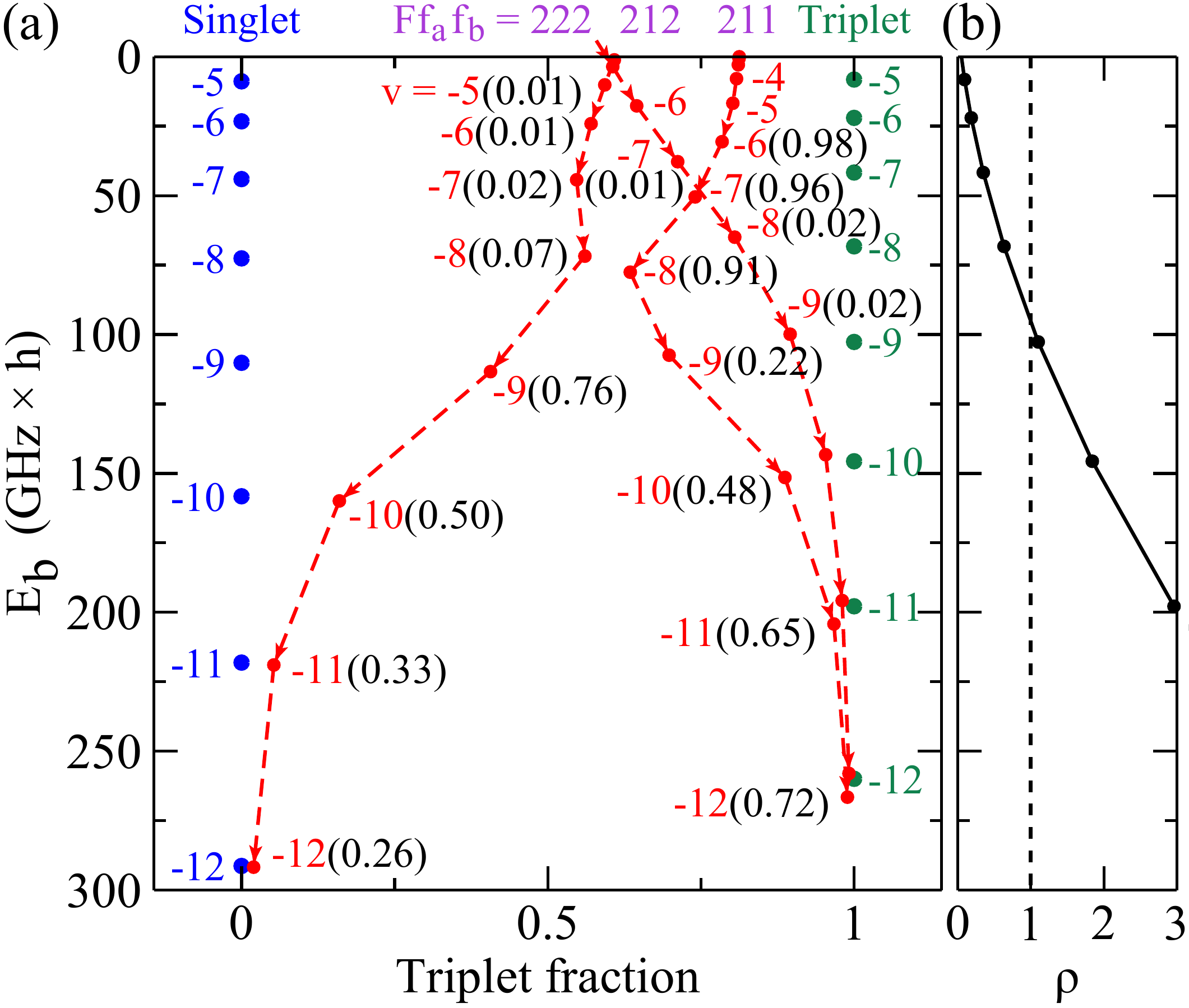}
	\caption{(a) Binding energy and spin character of weakly-bound $^{87}$Rb$_2$ molecules for various vibrational quantum numbers $\vib = -1$ to $-12$ at $B=0$ (red circles). Here, $L_R=0$, $F = 2$, and  $m_{F} = -2$. Zero energy corresponds to two separated atoms in state $f_a f_b = 1 1$. The spin character is given as the norm of the spin triplet component. Shown are the three spin families with $F = 2$ which correlate with the states $Ff_af_b = 222, 212, 211$ at the $E_b = 0$ threshold. Red dashed arrows are guides to the eye and indicate the change of each family's spin character with increasing binding energy. The norm of the 211 spin component is presented in parentheses. Blue and green circles are the bound state levels of the pure singlet and triplet potentials, respectively. The dissociation limit of these two potentials  is at $E=+8.543\:\textrm{GHz}\times h$ above $E_b=0$. (b) shows the ratio $\rho$ for various binding energies.}
	\label{SchematicRb87}
\end{figure}

\begin{figure}[htbp]
	\renewcommand{\thefigure}{S6}
	\includegraphics[width=\columnwidth]{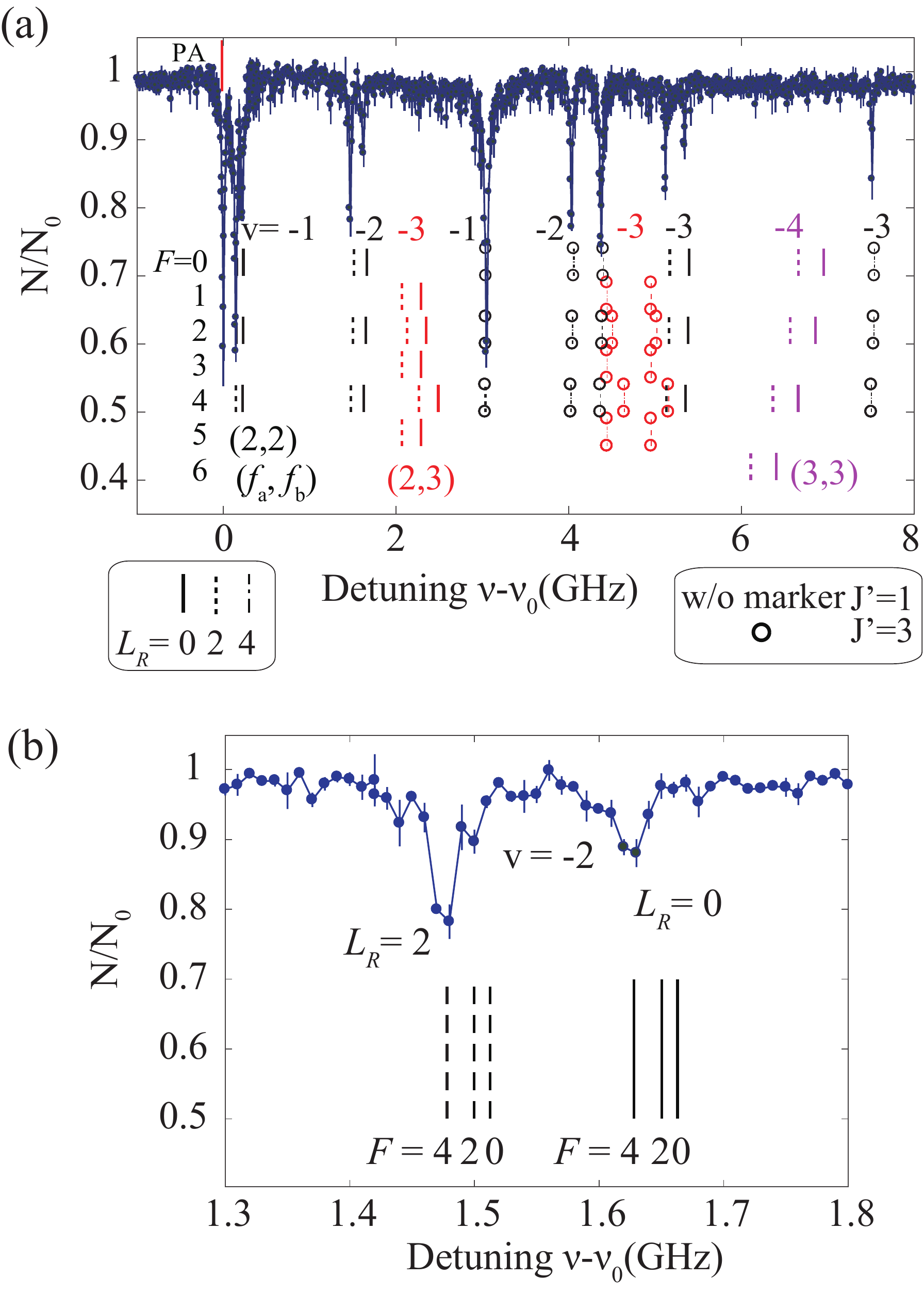}
	\caption{(a) REMPI spectrum of Fig.$\:$1(b) for small REMPI laser detunings $\nu-\nu_0$ between about zero and $8\:$GHz (path (S) via intermediate state $(2)^1\Sigma_u^+$, $\vib'=22$). Vertical lines represent calculated resonance positions for product molecules of spin families with $f_af_b=22$ (black), $f_af_b=23$ (red) and $f_af_b=33$ (purple). Each of the bound states with different $F$ (from 0 to 6) is separately plotted in the vertical direction. Negative numbers above the calculated bound state positions indicate vibrational quantum numbers. (b) Further zoom into the spectrum for REMPI laser detunings $\nu-\nu_0$ between $1.3$ and $1.8\:$GHz. Here, the black vertical lines correspond to the predicted resonance positions for product molecules characterized by $F, f_a=2, f_b=2$ with $F=0,2,4$.}
	\label{Fig:checkFstates}
\end{figure}

In order to gain more insight on how three-body recombination occurs within the hyperspherical representation we introduced an artificial hyperradial hard-wall at $R = R_{\rm hw}$ and performed recombination calculations for various values of $R_{\rm hw}$. The hard-wall prevents any flux into the range $R<R_{\rm hw}$ and therefore reactions in this range are not possible. By analyzing how the total and partial reaction rates change as a function of $R_{\rm hw}$ one can infer at what hyperradii $R$ reactions dominantly occur. Figure~\ref{L3Rhw} shows the total rate $L_{3}$ (black solid line) and the partial rates for some of the currently observed molecular states, $(\vib,L_R)$ = (-1,2), (-1,0), (-2,6), (-2,4), ({-2},2) and (-2,0). This particular calculation was performed with a two-body interaction model (Eq.$\:$\ref{vpot}) supporting 5 $s$-wave bound states in order to reduce some numerical instabilities associated with the hard-wall interaction. The figure clearly indicates that recombination is most likely to occur at  $R\approx1.5r_{\rm vdW}$. For  $R_{\rm hw} <1.5r_{\rm vdW}$ the partial rates are affected by interference effects related to the different collision pathways three atoms can follow to form a molecule \cite{dincao2018JPB}. Interference effects can be seen as the fast oscillations on the partial rates in Fig.~\ref{L3Rhw}. For $R>1.5r_{\rm vdW}$ partial rates are increasingly suppressed due to the decrease of the inelastic couplings with $R$ \cite{dincao2005PRA}. The peaking of the reaction rates at around $R\approx1.5 \, r_{\rm vdW}$
in Fig.~\ref{L3Rhw} can be explained as a consequence of a repulsive barrier in the hyperspherical effective potential of the entrance channel at $R\approx1.7r_{\rm vdW}$. Such a barrier was first identified in Ref.~\cite{Wang2012prl} where it occurred at $R\approx2r_{\rm vdW}$ for $a=\pm\infty$.

We now analyze the geometric characteristics of the three-atom system near the region where recombination is most likely to occur in order to identify which type of coupling is dominating the inelastic transitions. Figure \ref{ChanFun} shows the hyperangular probability density at $R = 1.593r_{\rm vdW}$ in terms of the hyperspherical channel functions $\Phi_\alpha$ and
hyperangles $\theta$ and $\varphi$ as: $|\Phi_\alpha(R;\theta,\varphi)|^2\sin(2\theta)$ (see 
Ref.~\cite{Wang2012prl}). Each point in the $\theta$-$\varphi$ hyperangular plane corresponds to a specific geometry of the three-atom system. In particular, the regions within the dashed half-circles represent geometries  where two of the atoms are found at a distance
$r_{ab} \le r_{\rm vdW} $ and $r_{ab} \le r_{\rm hf}$, respectively, and the third atom remains at distances $r_{ca},r_{cb} > r_{\rm hf}$ (i.e., $ r_{ci} \ge 0.71 r_{\rm vdW}$ and $ r_{ci} \ge 1.2 r_{\rm vdW}$, respectively). In Fig.~\ref{ChanFun}(a) we show the probability density for the initial collision channel, which indicates that the most likely configuration is where all three atoms are located at distances typically larger than $r_{\rm hf}$. In contrast, the  probability density for the target molecular states is pronounced  inside the small circle. As an example, Fig.$\:$\ref{ChanFun}(b) shows the probability density for $(\vib,L_R) = (-4,0)$. In order to produce this state, the two atoms forming the bound state obviously need to be closer than $r_{\rm hf}$ and the third atom is at interparticle distances $r_{ci} > r_{\rm hf}$.
For a spin flip to happen during the reaction, the two atoms forming the molecule must be at a distance $r_{ab} < r_{\rm hf}$ and the wavefunctions of either the two-body scattering state  or the two-body target bound state must have a sizeable spin admixture. In the range of binding energies discussed here for $^{85}$Rb these admixtures are on the \%-level and therefore quite small. The admixtures are not shown in Fig.~\ref{ChanFun}.

\subsection{Characteristics of collisions at large distance}

The interaction between two Rb atoms is effectively spin-independent for interparticle distances $r_{ij}  \equiv  | \vec{r}_i - \vec{r}_j | \gtrsim  0.25 r_{\rm vdW}$. This is  because at these distances the exchange splitting  between the electronic singlet and triplet ground state potentials is smaller than the atomic hyperfine splitting $E_{\rm hf} = 3.04\:\textrm{GHz}\times h$. As a consequence the atomic hyperfine states are only weakly perturbed by the exchange interaction and therefore essentially remain eigenstates in this realm. In other words the exchange interaction is not strong enough to decouple the atomic hyperfine spins and flip them.

Next, we consider the low-energy collision of two $f = 2, m_{f} = -2$ Rb atoms, corresponding to the channel $F f_a f_b = 422$. At vanishing magnetic field and neglecting dipolar relaxation processes, $F$ and $m_F$ are  conserved throughout the collision. At a close enough distance $< 0.25 r_{\rm vdW}$ spin exchange interaction, however, can admix the spin states $F f_a f_b = 423,433$ (For an overview of all spin channels of two colliding ground state $^{85}$Rb atoms see \cite{Koehler2006}.). Interestingly, this admixture is restricted to distances $r_{ab} < r_{\rm hf}\approx0.6r_{\rm vdW}$ due to energetic closure of the $f_a f_b=23, 33$ scattering channels at larger distances. For $m_F = -4$, the channel $F f_a f_b = 422$ is the lowest in energy. The hyperfine distance $r_{\rm hf}=(C_6/E_{\rm hf})^{1/6}$ is the distance at which the potential curve of channel $f_a f_b=23$ crosses the collision energy (which approximately corresponds to the asymptotic energy of the 422 channel). This crossing is also relevant in another regard. It effectively limits the size of molecular bound states of the spin channels $f_a f_b=23, 33$ which are energetically located below the $F f_a f_b = 422$ entrance channel to be $ \lesssim r_{\rm hf}\approx0.6r_{\rm vdW}$. 

We now consider a three-body collision  of three $f = 2, m_{f} = -2$ atoms and assume the hyperradius to be $R > 1.1 r_{\rm vdW}$. Using the expression $R^2 = (\vec{r}_b - \vec{r}_a )^2 /d^2 + d^2 (\vec{r}_c - (\vec{r}_a + \vec{r}_b )/2  )^2 $, where $\vec{r}_i$ is the location of particle $i$ and $d^2 = 2/\sqrt{3}$ \cite{dincao2018JPB}, we calculate that for $R > 1.1 r_{\rm vdW} $ at least one of the three atoms, say $c$,   has distances $r_{ca}$, $r_{cb} >  0.6r_{\rm vdW}$ from both other atoms. This is outside the distance range 
$ r_{ca}$, $r_{cb} < r_{\rm hf}\approx0.6r_{\rm vdW}$ for atom $c$ to exhibit a spin admixture. Therefore, atom $c$ should have a pure spin state $f = 2, m_{f} = -2$, even if it had possibly undergone spin exchange interaction in a previous collision with one of the other two atoms. Reciprocally, we can conclude that the two other colliding atoms $a$ and $b$
must be in the collision channel $F f_a f_b = 422$. Thus, at a hyperradius $R > 1.1 r_{\rm vdW} $ the three-body system can be effectively decomposed (with respect to spin) into a two-body collision of two $f = 2, m_f = -2$ atoms and a third atom which is spin-wise only a spectator. 

\subsection{Spin composition of $^{87}$Rb$_2$ molecules}

Figure \ref{SchematicRb87} shows the binding energy and the spin composition of weakly-bound $^{87}$Rb$_2$ molecules. For $^{87}$Rb$_2$ the ratio $\rho$ is given by $\rho = |E_\mathrm{T}(\vib)-E_\mathrm{S}(\vib)| / E_{\mathrm{hf}} $. It becomes $ > 1$ for binding energies $E_b \gtrsim 100\:\text{GHz}\times h $. Around $\rho = 1$  a major change in the spin composition of the spin families occurs.

\subsection{Tests of the spin propensity rule}

In the spectra of Fig.$\:$1 and Fig.$\:$\ref{fig6} we showed that our observed lines match with  the calculated positions of molecular states with $Ff_af_b = 422$. For the sake of completeness we show here with a few examples, that our data do not match with the calculated spectra of other spin families. Figure \ref{Fig:checkFstates}(a) shows a section of the spectrum in Fig.$\:$1(b) together with calculated positions for product molecules with $f_af_b=23$ and $33$. The positions are indicated by vertical red and purple lines, respectively. Each of the bound states with different $F$ (from 0 to 6) is plotted separately in vertical direction. Clearly, there is no convincing match between these and the experimental lines within the detection limit. The observed lines only match calculated resonance frequency positions for molecules with $Ff_af_b=422$. In Fig.$\:$\ref{Fig:checkFstates}(b) we zoom in further into the spectrum shown in Fig.$\:$\ref{Fig:checkFstates}(a) and check for molecules with $Ff_af_b = 022$ and $Ff_af_b = 222$. Again, the observed signals only match in a convincing way for $Ff_af_b = 422$.

\end{document}